# Advanced Load Shedding for Integrated Power and Energy Systems

Bang L. H. Nguyen
Electrical and Computer
Engineering Department
Clarkson University
Potsdam, NY, USA
bangnguyen@ieee.org

Tuyen Vu
Electrical and Computer
Engineering Department
Clarkson University
Potsdam, NY, USA
tvu@clarkson.edu

Colin Ogilvie
Center for Advanced
Power Systems
Florida State University
Tallahassee, FL, USA
cmo15d@my.fsu.edu

Harsha Ravindra
Center for Advanced
Power Systems
Florida State University
Tallahassee, FL, USA
ravindra@caps.fsu.edu

Mark Stanovich
Center for Advanced
Power Systems
Florida State University
Tallahassee, FL, USA
stanovich@caps.fsu.edu

Karl Schoder
Center for Advanced
Power Systems
Florida State University
Tallahassee, FL, USA
schoder@caps.fsu.edu

Michael Steurer
Center for Advanced
Power Systems
Florida State University
Tallahassee, FL, USA
steurer@caps.fsu.edu

Charalambos Konstantinou
Center for Advanced
Power Systems
Florida State University
Tallahassee, FL, USA
ckonstantinou@fsu.edu

Herbert Ginn
Electrical Engineering
Department
University of South
Carolina, Columbia, SC
ginnhl@cec.sc.edu

Christian Schegan
Naval Surface Warfare
Center – Philadelphia
christian.schegan@navy.mil

*Abstract*— **This paper introduces an advanced load shedding algorithm to improve the operability performance of a medium voltage direct current (MVDC) integrated shipboard power and energy system. Outcomes are compared to a baseline algorithm while considering power generation contingency scenarios. The case study is conducted with a real-time, embedded algorithm implementation using a control hardware-in-the-loop (CHIL) setup.**

## I. INTRODUCTION

### A. Shipboard Power System Architectures

Toward the development of all-electric ships, shipboard power systems (SPS) become fundamental and attract more research efforts. To guide the technology development process of SPS, the Navy has released the Naval Power and Energy Systems Technology Development Roadmap [1]. This roadmap builds on the integrated power and energy systems (IPES) for improved capabilities and reduced life cycle costs. IPES incorporates power generation, distribution, conversion, and energy storage modules along with novel controls to provide loads with the desired power quality [2].

Both Medium Voltage Alternating Current (MVAC) and Medium Voltage Direct Current (MVDC) configurations can be employed in an IPES. While AC based options have traditionally dominated naval power systems, DC solutions promise to deliver improved flexibility and controllability. Thanks to developments in solid-state power electronics, MVDC components and systems could potentially offer the increased power density and efficiency compared to MVAC systems. Moreover, the need to incorporate multiple high-power, high-ramp rate load demands, power electronics based MVDC systems potentially provide the desired advantages [3]. MVDC systems can directly supply loads that operate at the DC voltage levels without extra power conditioning equipment. Also, the integration of Energy Storage Systems (ESS) can be more manageable than in MVAC, making the MVDC architectures the most suitable configurations for future electric ships [4].

In order to facilitate control developments, SPS architectures should be designed, evaluated, and validated in the early stage [5], [6]. To compare the merits of IPES architectures, the Electric Ship Research and Development Consortium (ESRDC) has developed and implemented a real-time simulation testbed for notional multi-zone SPS of 12 kV MVDC and 13.8 kV/60 Hz MVAC. These models are validated under several real-time simulation platforms [7].

### B. Controls Evaluation Framework (CEF)

To accomplish electric ship missions, an MVDC-based IPES requires thorough coordination of managements and controls from the system to the component level and referred to as Automated Power and Energy Management Systems (APEMS). In the early-stage design, various control and management algorithms are designed, tested, and compared. This process of controls evaluation is critical to guarantee the system's efficiency, resilience, and reliability. Through the controls evaluation process, the best system designs, operational procedures can be verified. At the core of this capability demonstration process is the Controller Hardware-in-Loop (CHIL) approach. Following the implementation of a real-time simulation testbed for notional multi-zone SPS [7], the ESRDC has initiated the controls evaluation framework

(CEF) to support development and evaluation of shipboard power system controls [8]. The framework was used here in testing the controls developed, and it facilitates the evaluation of benefits through a largely automated approach. Based on prior accomplishments [9], the ESRDC has established notional MVDC and MVAC SPS real-time simulation models with baseline controls via a CHIL based environment. The baseline controls provide a known, consistent, initial state for comparison and further evaluations. The evaluation process further supports incorporation of additional impacts such as communication conditions as described in [10].

This paper presents a case study for load shedding controls evaluation. An advanced load shedding algorithm was developed with an optimization-based method. The algorithm is evaluated and compared with a baseline algorithm, a rule-based load shedding method against a power generation contingency scenario under a ship mission using the developed metrics in the CEF.

The remaining parts of this paper are organized as follows: In Section II, baseline and advanced load shedding algorithms are presented. Section III details the results, evaluations, and comparisons of the studied load shedding versus the baseline one. Finally, Section IV concludes the paper with suggestions for future works.

## II. MISSION-BASED LOAD SHEDDING CONTROL

This section presents the problem formulation of load shedding controls. Load shedding is crucial to balance the supply and demand of power in the event of the loss of generation and the load demand is higher than the generation capacity. The load shedding strategy determines which loads should be curtailed while maximizing the overall benefit of the entire system. The continuity of service to loads to fulfill a mission is more important than, for example, the power loss or fuel-efficiency.

### A. Baseline Load Shedding Control

The baseline load shedding control is a stage-based load shedding scheme. The algorithm monitors individual generator loading ($L$). When the loading on the generator is greater than 1.0 pu, based on pre-assigned load priorities, loads are shed in stages. Loads are prioritized into three categories: non-vital, semi-vital, and vital loads. The load priorities can be changed based on a specific mission. Fig. 1 shows the algorithm with three stages: Stage 1) When generator loading is higher than 1.0 pu for more than 250*ms*, the non-vital loads are sequentially shut down; Stage 2) If the loading is still high for 2.5*s*, the semi-vital loads are curtailed; Stage 3) Vital loads are shed when the generator loading is still higher than 1.0 pu for over 5.0*s*. This baseline scheme does not consider the load priorities within each category.

### B. Advanced Mission-based Load Shedding Strategy

Load shedding optimization algorithm in SPS is to prioritize the mission and critical loads by maximizing the total weighted operability [11]. The operational status of a single load is designated by a variable $o_i(t) \in [0,1]$, where $o_i(t) = 0$ meaning it is shut down; $o_i(t) = 1$ means that it is fully operational. Some loads only have binary operation status or stepped status changes. Under a specific mission $\theta$, each load is valued by a weight $w_{i\theta}(t)$ indicating the significance (or contribution) of the load with respect to that mission. This weight can be changed under the mission period. The operability over the mission period can be computed as:

$$O_\theta(t) = \frac{\int_{t_0}^{t_m} \sum_{i=1}^n w_{i\theta}(t) o_i(t) dt}{\int_{t_0}^{t_m} \sum_{i=1}^n w_{i\theta}(t) o_{i\theta}^*(t) dt}, \quad (1)$$

where $o_{i\theta}^*(t)$ is the maximum operation status of load $i^{th}$ under mission $\theta$ at the time $t$. The operational statuses and weights can be considered unchanged within a sufficiently small period $\Delta t = t_m - t_0$. Therefore, the instantaneous mission-based weighted operability is calculated as follows:

$$O_\theta(t) = \frac{\sum_{i=1}^n w_{i\theta}(t) o_i(t)}{\sum_{i=1}^n w_{i\theta}(t) o_{i\theta}^*(t)}. \quad (2)$$

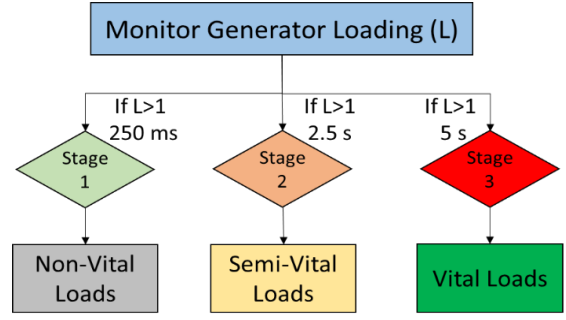

Fig. 1. Baseline load shedding control scheme.

This relative parameter reflects the percentage of load powers compared to the total demand powers at that time.

According to (2), under the normal operation of mission $\theta$, the mission-based operability would be 1.0 as all loads are operated following their demands. When an unexpected event occurs, causing power supply loss, some loads are selected to be cut for maximizing the weighted operability. Hence, the objective function of load shedding optimization can be formulated as:

$$\max_{o_i(t)} \sum_{i=1}^n w_{i\theta}(t) o_i(t). \quad (3)$$

The constraint of this optimization problem can be listed as follows:

*Supply-Demand Constraints*: The total load demand cannot exceed the generation power.

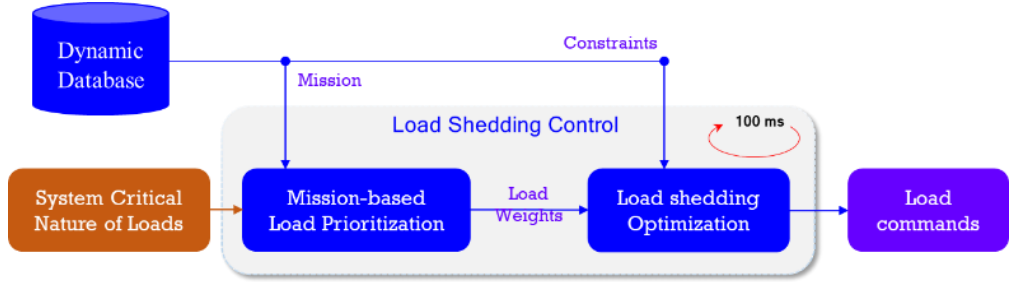

Fig. 2. Block diagram of mission-based load shedding control.

$$\sum_{i=1}^{n} P_{i\theta}^*(t) o_i(t) \leq \sum P_G(t) - \sum P_{Loss}(t), \quad (4)$$

where $P_{i\theta}^*(t) = o_{i\theta}^*(t) P_i(t)$ is the required power, $P_i(t)$ is the rated power, $\sum P_G(t)$ is the total generation power, and $\sum P_{Loss}(t)$ is the total power loss.

*Line Flow Constraints:* Although the generation power is adequate, damage may obstruct the power flow to the load location.

$$\sum_{i=1}^{m} P_{i\theta}^*(t) o_i(t) \leq P_{limit}^\delta, \quad (5)$$

where $P_{limit}^\delta$ is the limit power can flow into the area $\delta$, $m$ is the number of loads within this area.

*Physical Constraints*: Due to physical failures or attacks, some loads may be discarded directly from the operation.

$$o_j(t) = 0 \quad (6)$$

The mission-based load shedding control is summarized in Fig. 2, where the dynamic database block provides the mission information to prioritize loads. The constraints are also updated based on this dynamic database. The mission-based load prioritization will weigh the load values according to specific missions and the system critical nature of loads. Thereafter, the load shedding optimization can be solved, and load commands are released. This procedure is repeated every 100 ms in the real-time operation.

### III. CASES STUDIES AND RESULTS

This section presents the case studies and results of the evaluation of load shedding controls via the CEF. The MVDC SPS real-time simulation is conducted via the real-time simulator (RTS). The RTS allows users to apply different types of load profiles and run predefined test scenarios. Data logging hardware system capable of recording time-step level data for a long duration is utilized to record data. The advanced load shedding control algorithm is implemented in a real-time embedded industrial controller of National Instruments, named CompactRIO (NI-cRIO 9064). Inside NI-cRIO, the load weights are assigned, and load shedding optimization is solved. The cRIO controller receives mission modes, load demands, and the generator power conditions of the ship system from the RTS and sends the load shedding commands back to the ship system over a UDP-based communication protocol.

#### A. Automated Testing Procedure

Once the real-time MVDC shipboard power system model is established in the RTS, the automated testing procedure is executed. In this procedure, firstly, the data recording is configured. Secondly, the real-time SPS model is configured, and the mission profile is determined.

After the above configurations are completed, the advanced control in the NI-cRIO is started. The control algorithm waits for the feedback signal to execute load shedding commands. In the first test, the normal operation with load profile is maintained until 310s, when the main Power Generation Module 2 (MPGM2) trips offline. The total power generation capacity reduces to 60 MW. All data of the 600s testing period are collected for analysis.

The same test scenario is also repeated for the baseline load shedding. However, the baseline load shedding is internal to the RTS environment while the advanced load shedding is embedded in cRIO external of RTS.

#### B. Results Analysis

The total load power of baseline and advanced load shedding algorithms is compared with the load profile shown in Fig. 3. As can be seen, at 310s, due to an unplanned loss of MPGM2, the generation capacity ends up being lower than the total load demand, which initiates load shedding. Both the load shedding control algorithms curtail some loads.

Fig. 4 compares the operability of the baseline and optimal load shedding algorithms. With additional information on the load demand, the advanced load shedding improves the load operation with the reduced generation capacity of 60 MW from 310s-395s. In addition, the algorithm matches the load demand when the demand reduces to lower than 60 MW from 395s, whereas the baseline load shedding continues to curtail loads.

In detail, after tripping of MPGM2, the operability in baseline load shedding dropped significantly, while the advanced load shedding kept the operability higher than 0.95 and restored the operability to 1.0 when the load demand is smaller than the generation capacity. It is also noted that there are flickers in Fig. 4 because of the transients of the load demand. In detail, at the time step that the high-power load demand suddenly changes, the load power is physically delayed. This leads to a gap between the demand and actual power, which causes a flicker to occur in the calculated operability. The flickers can be eliminated in an improvement of load implementation in the RTS.

Regarding individual load groups performance, Fig. 5 and Fig. 6 show that all the AC non-vital loads and part of AC vital loads are cut in baseline load shedding. In contrast, the advanced load shedding retains the operation of those loads; however, it reduces the power of the propulsion motor modules (PMM), as shown in Fig. 7. Fig. 8. Fig. 9 show that the operation of critical loads such as IPNC and MW-class loads is maintained in both test cases as expected.

It is noted that the AC load center (ACLC) vital loads have the same weight as the PMM loads as shown in TABLE I. As PMM loads have higher rated power than ACLC, the reduction in PMM power instead of ACLC would improve the overall system operability. This explains why the advanced load shedding algorithm curtailed the PMM loads instead of ACLC loads.

## C. Evaluation

The load shedding algorithms are evaluated in terms of the computation time, the number of feedback signals, communication, and the operability as shown in TABLE II.

Since the baseline load shedding algorithm only assesses the generation loading signal to curtail loads, its computation time is negligible. In the advanced load shedding algorithm, an optimization problem must be solved with information of total generation capacity and current load demands; therefore, the computation time would be higher. In this case, the integer linear programming of 42 variables is solved, the computational time would be smaller than 50 *ms*. The achievable integral operability over the 10 minutes mission period of advanced load shedding is higher at 0.9973 while the integral operability of baseline load shedding is 0.829.

Regarding the feedback signals, the mission-based load shedding optimization requires information on the total generation capacity and the current load demand via UDP communication to solve for the load operation commands and improve operability. This is a trade-off since the baseline load shedding control only requires the internal generator loading signal to curtail loads.

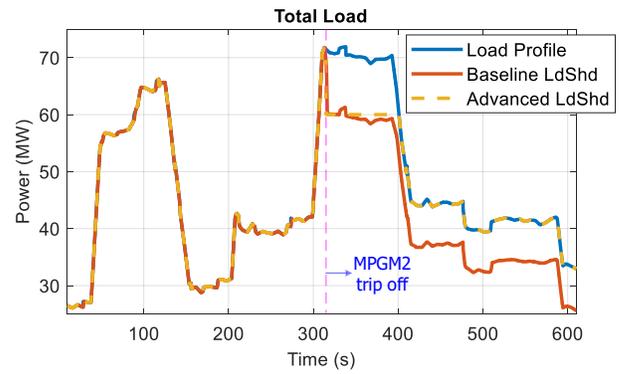

Fig. 3. Total load power.

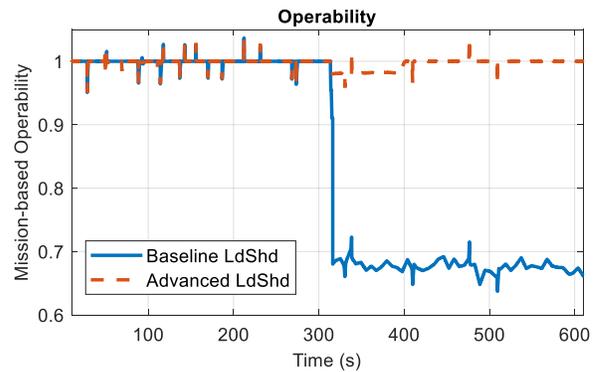

Fig. 4. System operability.

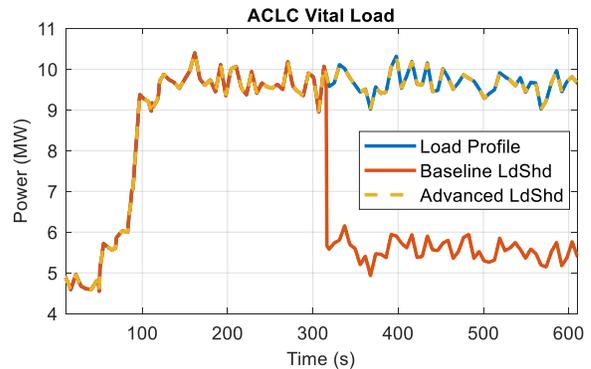

Fig. 5. Total ACLC vital loads.

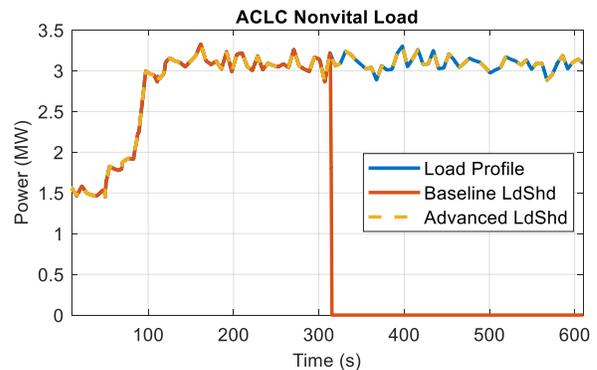

Fig. 6. Total ACLC non-vital loads.

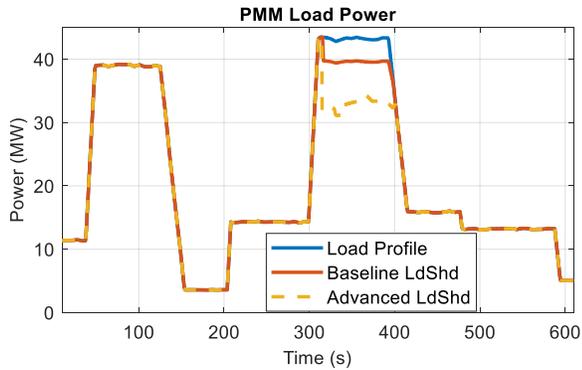

Fig. 7. The total PMM load power in the load profile, baseline, and advanced load shedding cases.

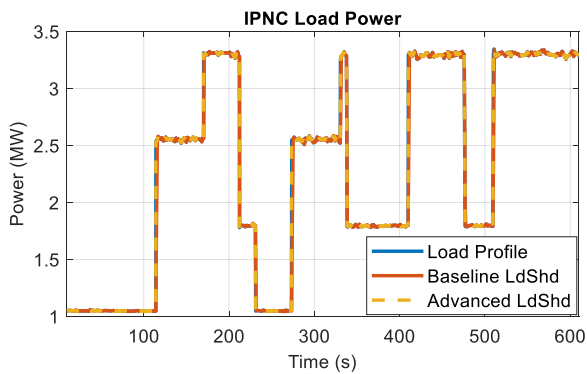

Fig. 8. The total IPNC load power in the load profile, baseline, and advanced load shedding cases.

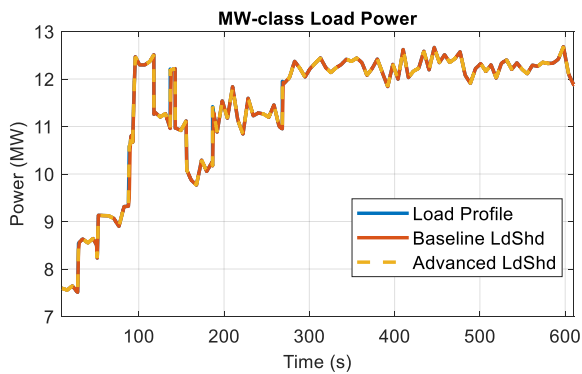

Fig. 9. The total MW-class load power in the load profile, baseline, and advanced load shedding cases.

## IV. CONCLUSION AND FUTURE WORKS

This paper presents a novel load shedding approach and its evaluation using a previously developed control evaluation framework. The case study provides a practical example for leveraging the advanced controls in a realistic, real-time based test bed environment, and both algorithm correctness and benefits have been evaluated. The simulation-based outcomes were analyzed and the evaluated using operability metrics to compare performance against a baseline load shedding approach. In the next evaluation phase, additional metrics will be considered including power ramp rates, computer network requirements (e.g., latencies, data loss), and distributed implementations of load shedding controls.

TABLE I. MISSON-BASED LOAD WEIGHTS

| Loads | Weights |
|---|---|
| ACLC Vital loads | 5 |
| ACLC Non-Vital loads | 2.5 |
| MW class loads | 8 |
| IPNC loads | 5 |
| PMM loads | 5 |

TABLE II. LOAD SHEDDING ALGORITHM EVALUATION

| Metrics | Baseline LdShd | Advanced LdShd |
|---|---|---|
| Computation time | Negligible | <50 ms |
| Communiation | Internal | External-UDP |
| Feedback signals | Generator Loading Signal | Generation Capacity, Load Demand |
| Operability | 0.829 | 0.9973 |

### ACKNOWLEDGEMENT

This material is based upon research supported by, or in part by, the U.S. Office of Naval Research under award number N00014-16-1-2956.